\documentclass[epj]{svjour}
\usepackage{latexsym}
\usepackage{graphicx}
\begin{document}
\newcommand{\goo}{\,\raisebox{-.5ex}{$\stackrel{>}{\scriptstyle\sim}$}\,}
\newcommand{\loo}{\,\raisebox{-.5ex}{$\stackrel{<}{\scriptstyle\sim}$}\,}
\title{Production of double strange hypernuclei and exotic nuclei in central Au+Au 
collisions at $\sqrt{s_{NN}}$=3 GeV}
%
\author{N.~Buyukcizmeci\inst{1} T.~Reichert\inst{2,3,4}, A.S.~Botvina\inst{2,3}, M.~Bleicher\inst{2,3,5}}

\institute{$^1$Department of Physics, Sel\c{c}uk University, 42079 Campus, 
Konya, T\"urkiye\\
$^2$Institut f\"ur Theoretische Physik, J.W. Goethe 
University, D-60438 Frankfurt am Main, Germany\\
$^3$ Helmholtz Research Academy Hesse for FAIR (HFHF), 
GSI Helmholtz Center, Campus Frankfurt, Max-von-Laue-Str. 12, 
60438 Frankfurt am Main, Germany\\
$^4$ Frankfurt Institute for Advanced Studies (FIAS), 
Ruth-Moufang-Str.1, D-60438 Frankfurt am Main, Germany\\
$^5$GSI Helmholtz Center for Heavy Ion Research, 
Planckstr.1, Darmstadt, Germany} 


%
%
%
\date{Received: date / Revised version: date}
%
\abstract{We extend the theoretical approach which includes the dynamical 
and statistical stages for the description of the nucleosynthesis 
in central collisions of relativistic ions. Previously, this approach was 
successfully applied to describe experimental data on both normal nuclei 
and single strange hypernuclei production in the GSI and RHIC-BES energy 
range. We predict the multiplicities of double strange hypernuclei up to 
$^4_{\Lambda \Lambda}$H and further intermediate mass nuclei up to $^8$Be 
for Au+Au central collisions at $\sqrt{s_{NN}}$=3 GeV, recently explored 
by the STAR experiments. These new nuclei can be 
identified by the measurement of the correlated particles coming after 
their decay. Such observations are a crucial test for the 
nucleosynthesis mechanism. 
\PACS{25.75.-q , 24.60.-k , 25.70.Pq , 21.65.+f }} 
\maketitle
\section{Introduction} 

The production of novel nuclei (including heavy and exotic ones) has 
always been an important and interesting topic in nuclear reactions. 
Typically, such processes include the interactions of multiple particles. 
A well known example of such a collective process is the compound nucleus 
formation and its subsequent decay 
\cite{Bohr1936,Bohr1939,Weis1937,Eric1960}, which takes place at low 
energies (i.e., essentially lower than the nuclear binding energy). 
The compound nucleus concept was very successful for the description of 
a large body of reactions (see, e.g., Refs.~\cite{SMM,Bot06} and 
references therein). Later on, a new collective process was established 
by studying the multifragmentation reactions: If a considerable amount 
of energy is deposited in the nucleus (more than 2--3 MeV per nucleon) 
such nuclear system can expand up to some freeze-out volume, and there 
the nuclei can be separated from uniform nuclear matter 
\cite{SMM,Bot06,Gros90}. This concept was confirmed with experimental 
and theoretical studies, and the parameters of the freeze-out volume 
were evaluated. Such a process takes place at subnuclear densities from 
0.1 to 0.3$\rho_0$ ($\rho_0 \approx 0.15$ fm$^{-3}$ being the ground 
state nuclear density). The temperatures of such systems are in the 
range from 5 to 8 MeV. These nuclear matter parameters correspond to the 
coexistence region in the nuclear liquid-gas type phase transition 
\cite{Sau76,Jaq84,Pan85,FASA}. Such matter at subnuclear density is the 
most suitable place for the formation of nuclei involving collective 
reactions of many nucleons: Because, 
1) at higher densities the formed nuclear clusters are destroyed by 
the interactions with other nucleons during the matter's expansion, 
2) at lower densities the collective nucleation processes are suppressed 
by too large distances between the nucleons. 
This reaction picture is consistent with state-of-the-art dynamical 
calculations describing the fragment formation also (see, e.g., 
Ref~\cite{Ono20,FRI,Aic20}). 

In the previous statistical description of the multifragmentation 
usually a single freeze-out source was considered in a single reaction 
event. For example, the source can be associated with highly excited 
projectile and target residues in peripheral nucleus-nucleus collisions 
\cite{SMM}. In this case the relativistic ion collisions open new 
possibilities to obtain hypernuclei, including multi-strange and exotic 
nuclei \cite{Bot16,Bot17,Bot13}, which are 
more difficult to produce in other reactions. 

During recent years central collisions of relativistic nuclei have been 
recognized an as important reaction channel leading to the production 
of novel (exotic) nuclei. A great variety of light complex nuclei can 
be formed in central heavy-ion reactions \cite{Gos77}. These studies 
were extended by involving the production of both normal nuclei and 
hypernuclei, including exotic nuclear species 
\cite{FOPI97,Neu03,Neu00,FOPI10,And11,ALICE,STAR,HADES}. Especially, 
the nucleosynthesis of strange nuclei can be studied in these processes 
in great detail. This opens a complementary route to better understand 
the nucleosynthesis in the early Universe. It is commonly accepted that 
light nuclei are mostly formed on later stages of the reaction from 
baryons which are primary produced 
\cite{FRI,Aic20,Gos77,Neu03,FOPI10,Ogu11,EOS,Gla22,Ton83}. 
Although other production mechanisms, like 
a direct thermal production, have been considered \cite{And11}. 

Recently we have proposed and validated a theoretical approach to explain 
the collective processes leading to the event-by-event production of a 
large variety of new nuclei in central nucleus collisions: We used the 
concept of local equilibrium with the formation of several statistical 
sources in a single reaction event \cite{Bot21,Bot22,Buy23}. The 
comparison with experimental data was very convincing \cite{Bot21,Bot22}. 
In addition, we have shown that the STAR experimental data on nuclei and 
hypernuclei production can also be described in the same way \cite{Buy23}. 
In this paper we continue our investigation with the STAR central 
collisions of Au+Au at $\sqrt{s_{NN}}$=3 GeV. We will demonstrate that 
besides normal and single strange hyper-nuclei many double strange 
hypernuclei can also be produced in this case. Also, further exotic and 
short-lived nuclei are predicted. Their experimental identification in 
the experiment will provide important information on the properties of 
hypermatter formed by the baryons produced during the initial dynamical 
stage. 

\section{Initialization of  baryons for the formation of nuclei in 
low-density matter}

The most practical way to describe these reactions is its subdivision 
into the two stages of different concept: (1) The dynamical interaction 
of incident nucleons and hadrons inside the nuclear matter leading  
to the formation of equilibrated nuclear systems, (2) the statistical 
fragmentation of such systems into individual nuclear fragments with 
the de-excitation of the hot fragments since they 
usually include excited states. Various transport models are currently 
used for the description of the dynamical stage of this reaction by 
involving binary hadron-hadron collisions at high energies. They take into 
account the individual particle interactions including the influence of 
the surrounding matter, the secondary interactions, and the decay of 
hadron resonances (e.g. \cite{Aic91,Ble99}). In this case the final 
particle distribution in the end of the dynamical part 
preserves important correlations between hadrons originating from the 
interactions in each event, which are ignored when we consider the 
final inclusive particle spectra only. Within these kinetic models it was 
established that many particles participate in these processes by the 
intensive rescattering leading to the final particle distributions which 
may look similar to statistical distributions with phase space 
domination. For example, in peripheral collisions the produced high 
energy particles leave the system and the remaining (spectator) nucleons 
form an excited system, a so called residue. We may expect 
that this system evolves toward a state which is mostly determined by 
the statistical properties of the excited nuclear matter. The system then 
expands by both a thermal pressure and the initial longitudinal motion 
to a low density and its multifragmentation leads to the production of 
various new (hyper-)nuclei (see, e.g., \cite{SMM,Ogu11,EOS,Bot07}). If 
the excitation energy of the residues is too low to enter the freeze-out 
state, then the nuclei formation proceeds via a compound nucleus \cite{SMM}. 
In the case when hyperons are captured by the residues, the final 
hypernuclei are produced as a result of the residue's decay 
\cite{Bot17,Bot13,lorente,Buy18,Bot15}. 

In central nucleus-nucleus collisions another physical picture can be 
realized: There are practically no spectator nucleons, and all produced 
baryons have a considerable kinetic energy. After the dynamical stage of 
the collision 
the system rapidly expands. At a time around $\sim$20--40 fm/c the baryon 
composition of this matter is established. During the expansion process 
some of the baryons may be located in the vicinity of each other with 
local subnuclear densities around $\sim$0.1$\rho_0$ 
($\rho_0 \approx 0.15$ fm$^{-3}$ being the ground state nuclear density). 
This nuclear matter density corresponds to the coexistence region in the 
nuclear liquid-gas type phase transition, and it is the proper place of 
the synthesis of new nuclei: The remaining attraction between baryons in 
the diluted matter can then lead to the formation of complex fragments. In 
the midrapidity kinematic region we expect a substantial production of 
hyperons, and subsequent hypernuclei production by hyperon capture. The 
formation process can be simulated as baryon attraction using potentials 
within the transport models \cite{FRI,Aic20,Gla22}, or within 
phenomenological coalescence models \cite{Neu00,Ton83,Bot15,Bot17a,Som19}. 
As we have demonstrated recently \cite{Bot21,Bot22}, the clusterization 
processes can be effectively described as the statistical formation of 
nuclei in the low density matter in local chemical equilibrium. We expect 
that such nucleation processes will mostly produce the light nuclei.

To describe the dynamical reaction part we use the transport 
model UrQMD \cite{Ble99,urqmd1}, which is adapted for these reactions.  
UrQMD is  quite successful in the description of a large body of 
experimental data on particle production \cite{Tom22,Tom23}. The 
produced particles can be located at various rapidities, however, the 
main part is concentrated in the midrapidity region. After a time of 
20--40 fm/c the strong interactions leading to the new particle 
formation cease and the system start to decouple. Such kind of a 
freeze-out is general for the transport approaches \cite{Bot11}. 
In this time-moment we consider the relative space positions and 
velocities of the baryons. We select the nuclear clusters 
according to the coordinates and velocities proximity, as was suggested 
in Refs.~\cite{Bot21,Bot15}, and we call it clusterization of baryons 
(CB). More details on the current UrQMD calculations for Au+Au collisions 
at $\sqrt{s_{NN}}$=3 GeV and the following CB procedure 
is provided in Ref.~\cite{Buy23}. Here we refrain from a full reiteration 
and mention only the most important CB parameter, $v_{c}$, which 
determine the maximum relative velocity of baryons inside the clusters 
respective to its center of mass velocity. As was established previously 
\cite{Bot21,Bot22} $v_{c}$ is within the range of $0.1c-0.25c$ (that is 
consistent with the Fermi motion). This parameter is mainly responsible 
for the cluster size and its excitation.  

The main goal of our paper is to investigate the production of 
double hypernuclei and other very rare nuclear species, which 
depends essentially on the phase space occupied by baryons. Therefore,  
in addition to UrQMD, we introduce another method to simulate the 
baryon velocity distributions after the initial reaction stage. Since the 
nuclear system expands after the collision in all directions it is possible 
to use a simplified phenomenological approach for the first reaction stage, 
in order to investigate the fragmentation mechanisms in detail. Here we use 
the phase space generation (PSG) method, in which we simulate an expanded 
nuclear matter state with stochastically distributed baryons: This is the 
isotropic generation of all baryons of the excited nuclear system according 
to a microcanonical momentum phase space distribution, with total momentum 
and energy conservation, corresponding to the one-particle approximation. 
It is assumed that all particles are in a large volume (at subnuclear 
densities) where they can still interact with others to populate the phase 
space uniformly.  Technically, this is done using the Monte-Carlo method 
applied previously in the microcanonical SMM and Fermi-break-up model 
\cite{SMM}, and taking into account the relativistic effects according to 
the relativistic connection between momentum $\vec{p}$, mass $m$, and kinetic 
energy of particles $E_{0}$, see Eq.~(\ref{relc}). In Eq.~(\ref{relc}) the sum 
is over all particles and we use units with $\hbar = c = 1$. 
\begin{equation} \label{relc}
\sum \sqrt {\vec{p}^2 + m^2} = E_0 + \sum m .
\end{equation}
The total kinetic energy available for the motion of baryons 
$E_{0}$ (the source energy) is the main parameter 
which can be adjusted to describe the energy accumulated in the system 
after the dynamical stage. We have checked that the PSG generated phase 
space distributions provide a reasonable assumption for the initial stage of 
the clusterization. 
Note, that this generation leads to the equilibrium over one-particle degrees 
of freedom, and it is not an equilibrium 
with respect to the nucleation process, i.e., it is not a global chemical 
equilibrium for all baryons in the system. 
In the PSG case we assume that the coordinates of the baryons are 
proportional to their velocities and strictly correlate with them as taking 
place in many explosive processes with strong radial flow. 

The main challenge in using the PSG is the correct estimate of the baryon 
content and the kinetic energy of the baryons in the system. At low 
(non-relativistic) beam energies, it would be equal to the total number 
of nucleons times the center of mass energy of the colliding nuclei, in 
central collisions. However, at relativistic energies many new hadrons 
are produced and the corresponding energy part should be subtracted, since 
this part is not participating in the nucleation process. To estimate this 
energy fraction we 
use the results obtained by the UrQMD calculations. Then in central Au+Au 
collisions at $\sqrt{s_{NN}}$=3 GeV we have in average 3.2 produced 
$\Lambda$ hyperons per event, and the average center-of-mass kinetic energy 
is 366 MeV per baryon. For this reason we take the $E_{0}$=366$A_0$ MeV 
for the PSG calculations, with $A_0$=394 including 3 $\Lambda$s, here  
$A_0$ denotes the total number of baryons considered. 

\begin{figure}[tbh] 
\includegraphics[width=85mm,height=120mm]{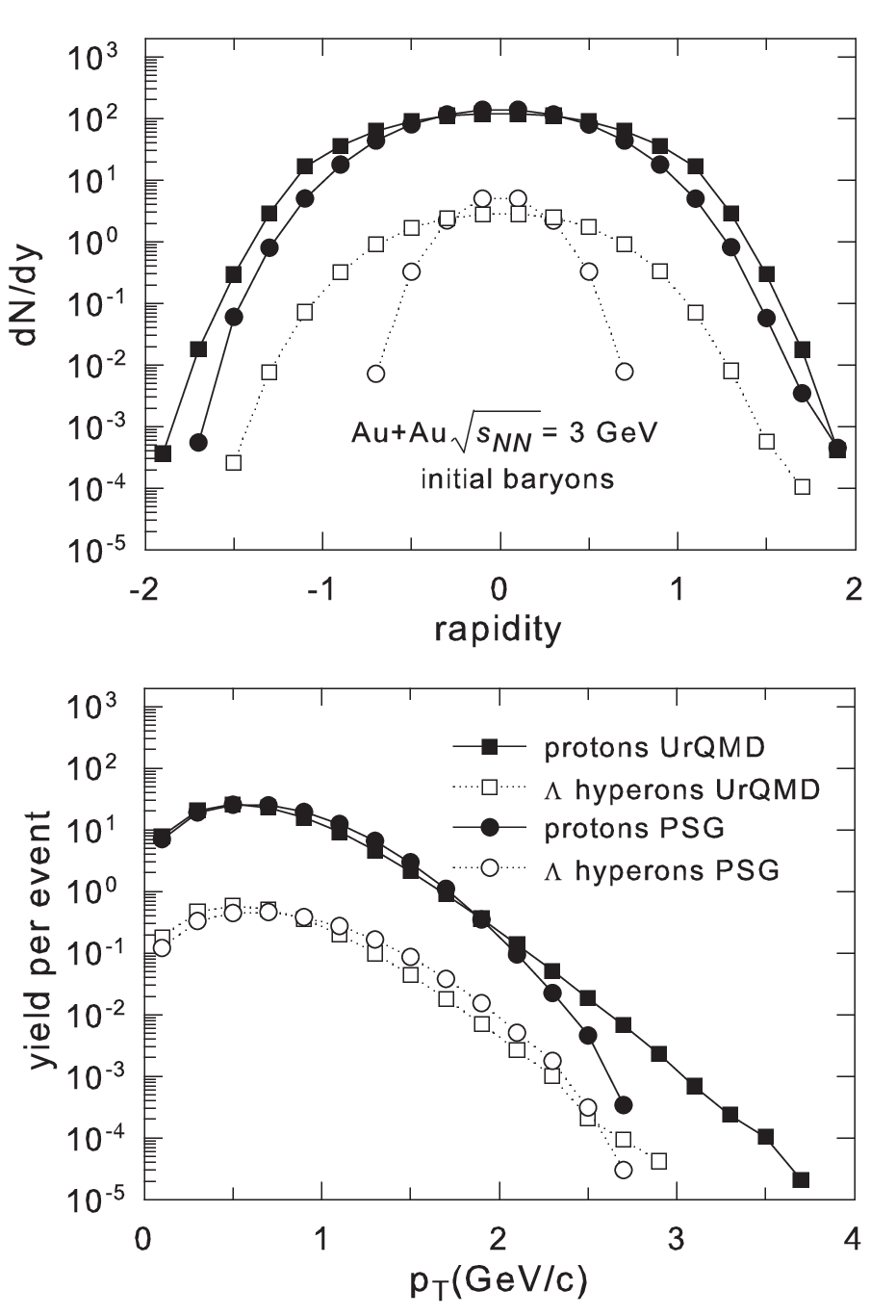} 
\caption{\small{ 
Total proton and $\Lambda$ distributions (per event) after UrQMD and PSG 
calculations of central gold collisions at center-of-mass energy of 
$\sqrt{s_{NN}}$=3 GeV. Top panel (a) - rapidity distributions. 
Bottom panel (b) -  transverse momenta distributions in the rapidity 
range $|y| < 0.5$. 
}} 
\label{fig1} 
\end{figure} 

In Fig.~\ref{fig1} we show the distributions of protons and Lambda hyperons after 
the UrQMD simulations in central Au~+~Au collisions (impact parameters 
$b \leq 3$~fm) at $\sqrt{s_{NN}}$=3 GeV center-of-mass energy. We show 
all nucleons and produced hyperons at the cut-off time $t=20$ fm/c, i.e., after 
the first hadron interaction inside the nuclei. We have checked that using later 
times changes our final results only slightly because the interaction rate is rapidly 
decreasing (see also discussion in Ref.~\cite{Buy23}). However, since the baryon 
coordinates may also be important for the cluster selection after UrQMD, in 
Figs.~\ref{fig2} and \ref{fig4} we'll present the comparison of 20 fm/c and 40 fm/c  
cut-off times. 

Also in Fig.~\ref{fig1} the corresponding PSG calculations are presented. Both 
UrQMD and PSG produce rather broad rapidity and transverse momentum 
distributions. However, the particle rapidities after PSG are more concentrated 
around midrapidity (especially for hyperons, since their masses are larger) than 
in the UrQMD simulations. This difference is due to the dynamical transparency, 
present in UrQMD since some fast nucleons after rescattering can go through the 
nuclear matter and interact later with the secondary hadrons to produce hyperons. 
This effect does not exist in the 
phase space simulation with PSG. Such a difference in baryon distributions gives 
us an opportunity to evaluate the effect of the uncertainty of the initial baryon 
production on the final production of nuclei. 

\section{Excited clusters at subnuclear density}

In both scenarios (UrQMD and PSG) we subdivide the diluted nuclear matter into 
multiple clusters with the CB procedure \cite{Bot15,Bot21,Bot22}. Since 
baryons move relative to each other 
inside the clusters these clusters present pieces of excited nuclear matter. The 
selected clusters have a baryon density around 
$\rho_c \approx \frac{1}{6} \rho_0$: It was established 
in the previous studies of the statistical multifragmentation process 
\cite{SMM,FASA,MMMC,Xi97,MSU,INDRA,TAMU,Vio01} that such a density 
is typical,  as a consequence of the liquid-gas type phase transition, for the 
fragmentation of matter at a low density freeze-out. The excitation energy of 
such clusters is calculated according to the method given in 
Refs.~\cite{Bot21,Bot22}. 
In the PSG case we have additionally 
taken into account the source energy ($E_0$) and momentum conservation in the 
system by adjusting these excitation energies. 
The following evolution of the clusters, including the formation of nuclei from 
the baryons, is now described in a statistical way. 
In our model this is described as the decay of hot clusters into light nuclei. 
For the description of this process we employ the Statistical 
Multifragmentation 
Model (SMM) which describes the production of normal nuclei very well, and 
use its generalization to the hypernuclear case 
(see Refs.~\cite{SMM,Ogu11,EOS,Bot07,lorente}). 

\begin{figure}[tbh] 
\includegraphics[width=8.5cm,height=13cm]{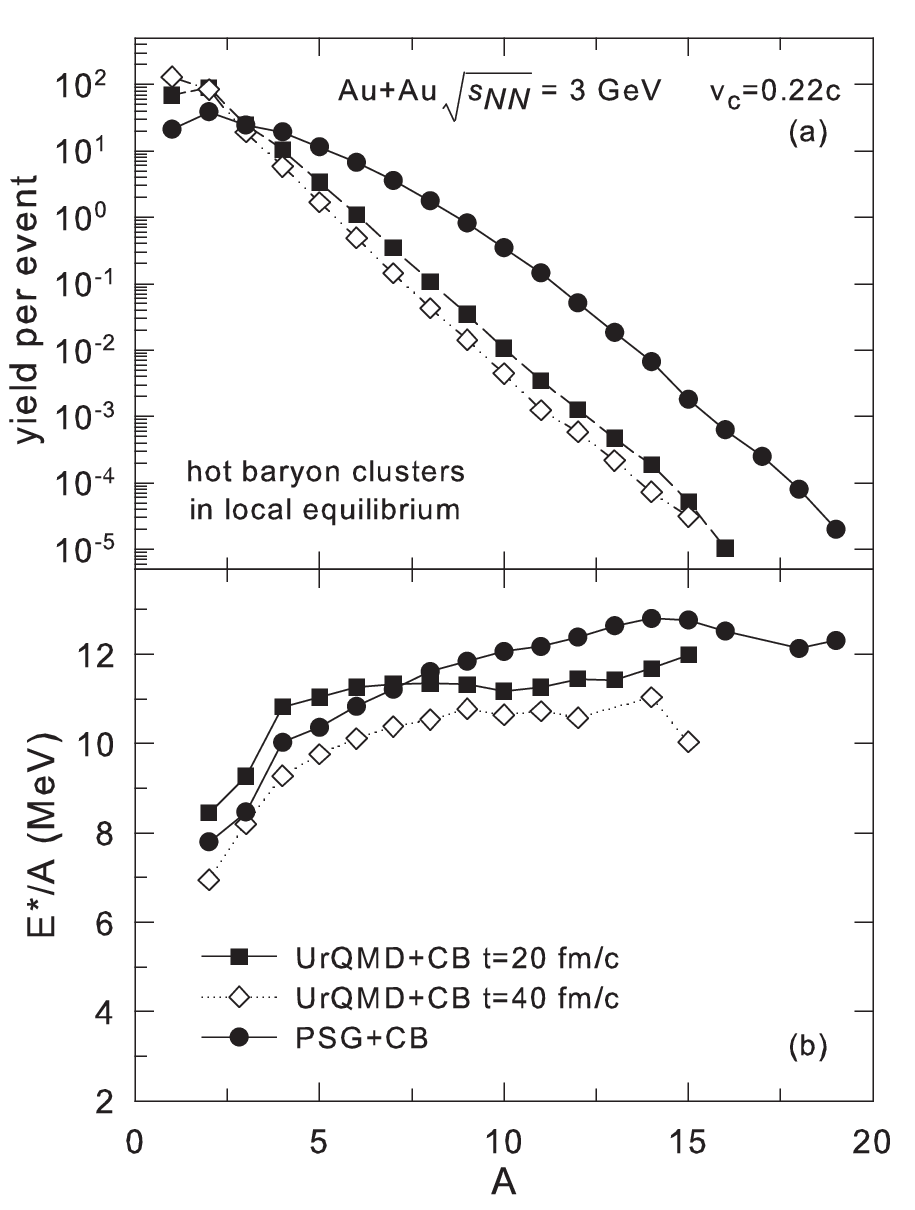} 
\caption{\small{ 
Comparison of distributions of local nuclear clusters (per event) 
formed after UrQMD dynamically produced baryons, and PSG statistical 
simulation for these baryons, by including the CB 
clusterization procedure which uses the selection of the  
baryons with the velocity and coordinate proximity. 
Top panel (a) - mass distributions of the cluster with the 
velocity parameter $v_c$=0.22$c$. 
Bottom panel (b) - average excitation energy of the clusters versus 
their mass number. The cut-off times for the UrQMD calculations 
are shown in the panels. 
}} 
\label{fig2} 
\end{figure} 

The results of the selection of the baryonic clusters within CB are shown in 
Fig.~\ref{fig2} after UrQMD and PSG. We  see that the cluster yields decrease nearly 
power-law like with their masses, and it is similar to a normal coalescence-like 
process. However, contrary to the coalescence, in our approach these clusters 
are excited nuclear systems in local chemical equilibrium. 
The important characteristic is the excitation energy of these clusters, which are 
presented in the bottom panel. Here we have used the previously established velocity 
parameter  $v_c$=0.22c. As discussed in Ref.~\cite{Buy23}, the heavier 
clusters can be formed with larger $v_c$'s. 
However, a large $v_c$ leads also to a high excitation energy, since the 
relative velocity spread of the baryons inside the cluster is higher. 
A slight time dependence of the cluster sizes for UrQMD is expected: 
It is obvious that the larger coordinate distances at later times between baryons 
decrease slightly the production of heavy clusters because of the increased spatial 
separation. One can see that the PSG leads to larger yields of massive clusters, 
and this is a very 
important effect related to the phase space population of the initial baryons. 
In UrQMD, the initial nucleons are separated from each other in the coordinate 
space in the very beginning, and after hadron-hadron collisions during the 
dynamical stage this separation is still preserved and influences the hot cluster 
selection. Therefore, the clusters in UrQMD are usually smaller than in PSG case, 
where we assume a direct correlation between the velocities and the coordinates of 
the nucleons. 
However, the excitation energies of the clusters remain very similar for 
all cases and depend mainly on $v_c$ (see \cite{Bot22,Buy23}). 

\begin{figure}[tbh] 
\includegraphics[width=8.5cm,height=13cm]{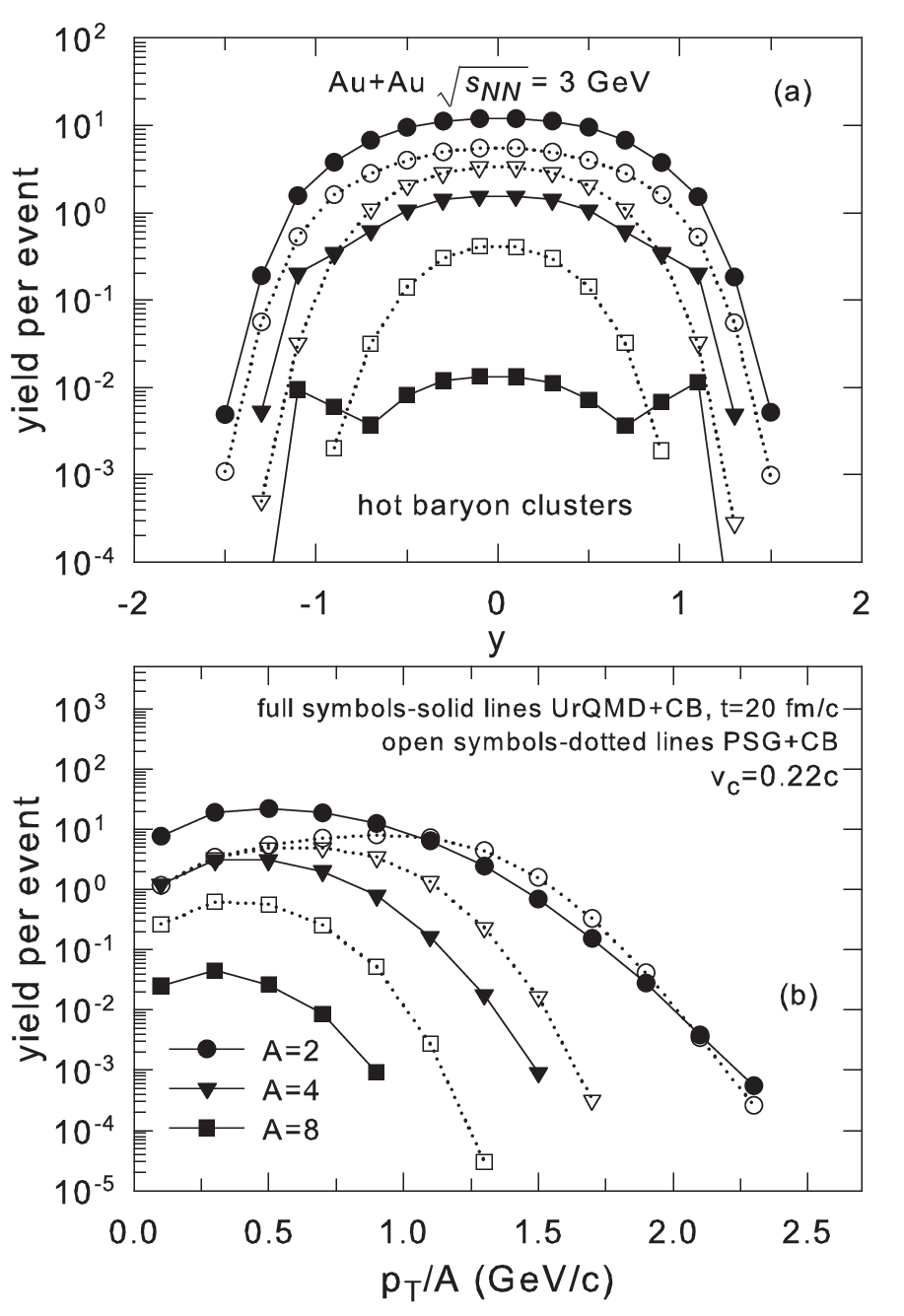} 
\caption{\small{ 
Yield distributions (per event) of the excited baryon clusters 
with mass numbers A=2 , A=4, and A=8
Top panel (a) - the rapidity ($y$ intervals of 0.2), bottom panel (b) - 
the transverse momentum per nucleon ($p_T / A$ intervals of 0.2 GeV/c). 
The calculations are UrQMD+CB (full symbols) and PSG+CB (empty symbols). 
}} 
\label{fig3} 
\end{figure} 

The kinematic characteristics of the primary excited clusters (rapidity and 
transverse momenta in the center-of-mass system) 
are depicted in  Fig.~\ref{fig3}, for $v_c$=0.22$c$. 
To evaluate the mass effect we have considered 3 groups of clusters 
with mass numbers $A$ equal to 2, 4, and 8. 
One can see that all distributions are quite broad because of the wide momentum 
distribution of the initial baryons in both UrQMD and PSG cases. The yields 
of massive clusters are low, that is consistent with Fig.~\ref{fig2}. Still 
there is a difference between UrQMD and PSG: The rapidity distributions of 
massive clusters (see A=8) in the UrQMD case can be partly enhanced by the 
spectator-like nucleons and be slightly increased towards the projectile and 
target rapidities. 

\section{Production of unstable nuclei and hypernuclei}

For further detailed information about the production of new normal nuclei 
after the decay of such excited clusters we refer to Refs.~\cite{Bot21,Bot22}. 
For example, the products of the cluster decay will preserve the kinematic 
characteristics (per nucleon) corresponding to the dynamically produced baryons. 
The isotope production can be explained by assuming the local chemical equilibrium. 
Many new exotic nuclei can be formed, and the specific correlations of hadrons 
are the best way to distinguish this reaction mechanism from normal coalescence 
or direct thermal production. 
Let us now address a very interesting phenomenon of the formation of 
short-lived nuclear states and hypernuclei which can be observed in experiments 
mainly via measuring particle correlations. We make a new suggestion concerning 
information which can be extracted from the extensive STAR experiment of 
Au~+~Au central collisions at $\sqrt{s_{NN}}$=3 GeV. In Ref.~\cite{Buy23} we 
have demonstrated that the fragment production in these central collisions can 
be described within our hybrid approach. 

\begin{figure}[tbh] 
\includegraphics[width=8.5cm,height=13cm]{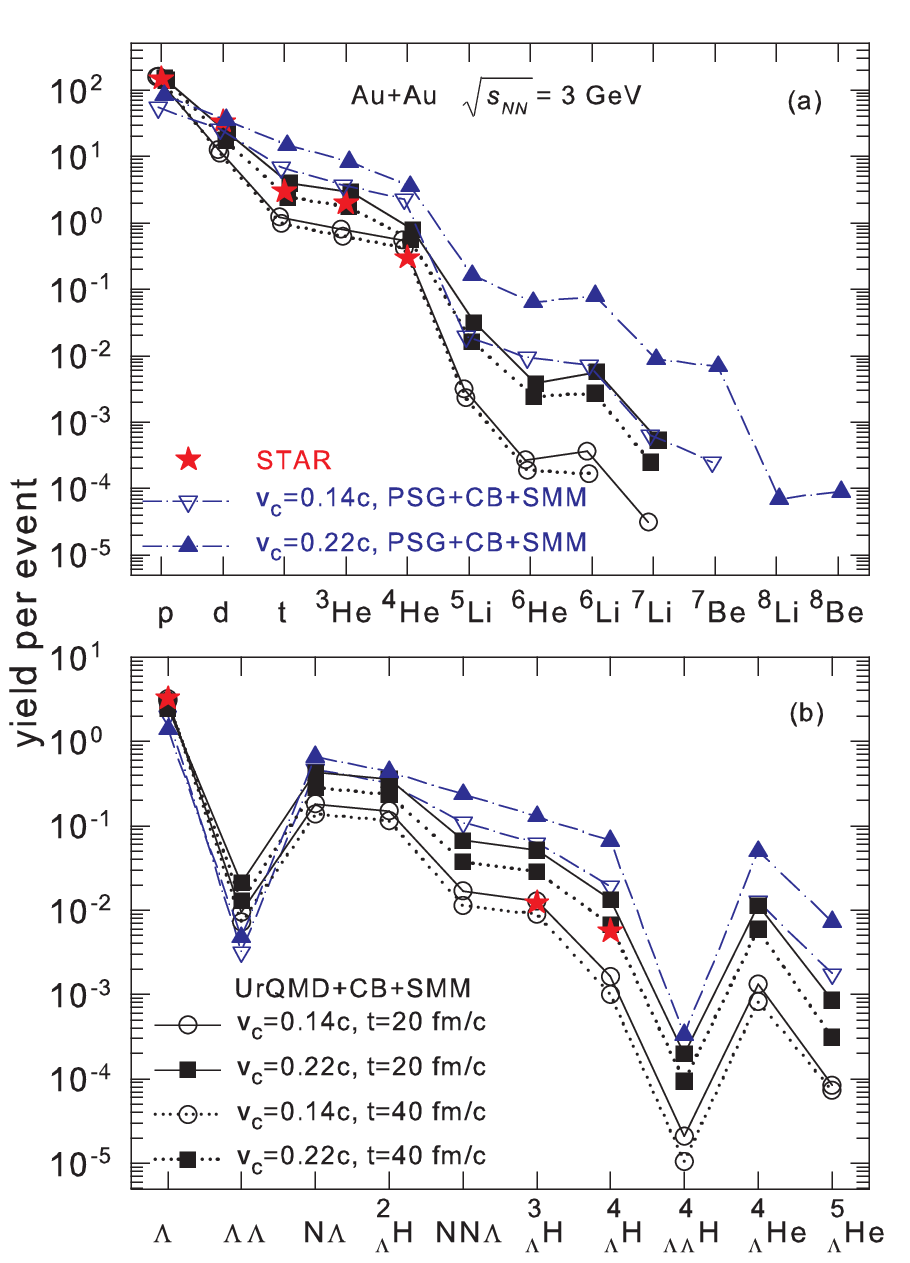} 
\caption{\small{ 
(Color online.) 
Comparison of the calculations of final nuclei production including UrQMD 
(black symbols) and PSG (blue symbols), formation of excited local thermalized 
clusters (CB) and de-excitation of these clusters (SMM). The STAR 
experimental data in central collisions (red symbols) are presented too.
Top panel (a) - total yields of normal nuclei. 
Bottom panel (b) -  total yields of hypernuclei in central collisions. 
Notations for nuclei, and the used UrQMD time and $v_c$ parameters are 
shown in the panels. 
}} 
\label{fig4} 
\end{figure} 

In Fig.~\ref{fig4} we show the final state yields of selected light nuclei and 
hypernuclei (normalized per event) obtained in central collisions. We compare the 
cold nuclei after the formation and decay chain UrQMD+CB+SMM and PSG+CB+SMM 
calculations. For convenience we show  also the published STAR experimental data 
\cite{STAR,STAR1}. As previously established, one generally reaches a good 
description of the data for normal fragments with the mass numbers range $A=2-4$ 
by using for UrQMD the velocity parameter $v_c$=0.22$c$, which corresponds to the 
hot cluster excitation energies around 
10 MeV per nucleon  (see Fig.~\ref{fig2} and the discussion in Ref.~\cite{Buy23}). 
For PSG case this parameter can be lower since the primary baryon clusters are 
bigger. However, both parameter sets are consistent with the parameter values 
extracted from the analyses of FOPI experimental data \cite{Bot21,Bot22}. 

The yields manifest the expected power-law decreasing with $A$. Nevertheless, we 
predict a substantial amount of nuclei with $A \geq 4 $. Such nuclei are very 
important for the understanding of the nucleation mechanism. For this reason, first, 
let us consider normal nuclei (top panel). We predict a relatively large yields 
of unstable $^{5}$Li nuclei (in the excited level of 16.8 MeV), which decays mainly 
into proton and $^{4}$He \cite{Ajz1955,Ajz1959} after a time of around 10$^{3}$ fm/c. 
The corresponding correlations can be observed in experiment, since the decay 
time of this state is substantially longer than its formation time in this 
multifragmentation reaction, which is less than 100 fm/c. One also observes 
considerable yields of stable and long-lived Li and Be nuclei. The measurement 
of such massive nuclei in high energy reaction in the midrapidity zone would an 
important confirmation of the local chemical equilibrium process. It is very difficult 
to explain the production of nuclei with A$>$4 within a simplistic coalescence 
model or within a global equilibrium hypothesis (with very high temperature). 
We suggest to measure correlations of particles coming from decays 
of unstable massive nuclei. For example, $^{8}$Be nuclei can decay into two $^{4}$He 
nuclei after a prolonged time of  10$^{-16}$ sec. We predict a substantial 
yield of $^{8}$Be (around 10$^{-4}$ -- 10$^{-5}$ per central event) by using PSG 
nucleon distributions. The experimental observation of such a correlation would 
indicate that nuclei with A=6--7 should be produced too. Within the present 
version of UrQMD we have estimated that the probability to produce these nuclei is 
very small ($\loo$ 10$^{-6}$). Because with the baryon distributions from UrQMD one 
obtains few large primary hot 
clusters. However, this probability may be higher within other considerations of the 
dynamical stage by using other model parameters or other dynamical models 
\cite{Bot21}. For all initial baryon distributions one can use the same statistical 
prescription describing the nuclei 
formation from neighbour baryons. A natural way for the identification of the 
reaction mechanism is the experimental observation of the particle correlation 
after decays of primary hot clusters, 
unstable nuclei and hypernuclei \cite{Bot21}. 
We emphasize especially that for such many-body phenomena the correlations 
provide the most effective method to unveil the real formation process 
among many alternatives. 

In relativistic collisions the yields of hypernuclei are very abundant, and we 
demonstrate it in the bottom panel of Fig.~\ref{fig4}. It is instructive to 
extend the nuclear chart into the strangeness sector, because hypernuclei can 
provide information on the hyperon interaction in nuclear matter, and on the 
properties of hypermatter, that is also important in astrophysical environment 
(see, e.g., Refs.~\cite{wakai1,Bando,Scha93,Scha96,japan,gal16,hiy18}). 
In addition, the suggested fragment formation mechanism realized in heavy ion 
collisions allows for investigating hypermatter at subnuclear densities. 
The production of hypernuclei as a result of decay of excited nuclear 
hyper-clusters was first addressed in peripheral ion collisions and hadron 
reactions \cite{Bot12}. In the reaction under consideration, see 
Ref.~\cite{Buy23}, we have discussed the production of well known 
light single strange hypernuclei. Here we present also the predictions  for 
exotic/unknown single strange hypernuclei, in particular, N$\Lambda$, 
$^{2}_{\Lambda}$H, and NN$\Lambda$ hypernuclei. It is interesting 
that the ground state hypernuclei with A=2 were not observed in 
the traditional hyper-nuclear experiments by using hadron/lepton beams. 
To this aim relativistic ion collisions provide new opportunities. In our 
theoretical approach we are able to consider both deeply-bound and short-lived 
excited states of these nuclei, if they exist. For a first rough estimate we have 
assumed that such nuclei have a very low hyperon binding energy 
of 0.05 MeV. It is encouraging 
that NN$\Lambda$ hypernucleus was reported in Ref.~\cite{Rap2013}. 
Therefore, it would be interesting to verify their existence in new 
experiments with relativistic ions. 

In this work we have increased the statistics of the UrQMD calculations 
for central collisions up to 96000 events, and we point at the possibility 
to produce double hypernuclei. One of the open puzzles of the hypernuclear physics 
is still the H-dibaryons, i.e., the bound $\Lambda\Lambda$ systems, which 
have been discussed for many years 
(see, e.g., Refs.~\cite{Bot12,Jaf77,Ste12,Gal24}). 
In the present calculations we have assumed a zero binding for the 
ground state and still obtain a quite large amount of $\Lambda\Lambda$, 
around 10$^{-2}$ per event. Strange H-dibaryons were not observed in 
experiments up to now. 
The reason may be that their lifetime is too short. Nevertheless, they may be seen 
in the correlation measurement by decay products in modern experiments. 
If we assume that there exists a deeply-bound state or relatively long-lived 
excited state then its yield may be several times higher or lower. However, 
it will be sufficiently high to be measured by particle correlations 
according to our predictions. It is further very 
instructive to measure $^4_{\Lambda\Lambda}$H hypernuclei in the well 
known ground-state. We predict these nuclei at the level of 
10$^{-5}$ -- 10$^{-4}$ for all calculation scenarios. A possible many-particle 
correlation to detect these hypernuclei might be  
$^4_{\Lambda\Lambda}$H$\rightarrow\pi ^{-}$+$^4_{\Lambda}$He , and 
then $^4_{\Lambda}$He$\rightarrow\pi ^{-}$+p+$^3$He. 
There was suggested recently that the double strange hypernuclei consisting 
of two $\Lambda$s and two neutrons (i.e., $^4_{\Lambda\Lambda}$n) may exist 
\cite{Ble19}. We have estimated from our calculation that, by assuming their 
binding energy in a low limit of 0.05 MeV, their yield would 
be several times smaller than the yield of $^4_{\Lambda\Lambda}$H.  
The observation of double strange hypernuclei and the comparison with the yields 
of single strange hypernuclei and normal nuclei will allow to investigate 
hyper-matter and hyperon-hyperon interaction at 
subnuclear density. This is a direct consequences of the statistical regularities 
of the production mechanism at the nucleation stage \cite{Buy18}. 

\section{Conclusions}

We have applied a new theoretical approach to explain the yields of 
conventional light nuclei (including exotic/unstable ones) 
and single/double strange hypernuclei which can be 
measured in relativistic ion experiments in central nucleus-nucleus collisions. 
Our approach combines 1) the adequate dynamical and statistical models to 
find the distributions of baryons produced in the first reaction stage, 
2) the formation of intermediate local sources in chemical equilibrium 
(excited multi-baryonic clusters) at subnuclear density, and 3) the 
description of the nucleation process inside these sources as their 
statistical decay. A special theoretical development of this work is 
the suggestion to use PSG method for the initial baryon distributions with 
general characteristics of nuclear matter taken from UrQMD calculations. 
This provides a chance to evaluate the effect of the variation of the 
initial baryon distributions on the nucleosynthesis. 
As was shown previously \cite{Bot21,Bot22} our approach 
can be successfully used to analyze the production of non-strange nuclei. 
We have demonstrated that the STAR experimental data concerning the production 
of nuclei and hypernuclei at a high collision energy can also be 
described within this approach by using similar parameters for the local 
sources. This may indicate a universal character of the nucleation process 
in rapidly expanding nuclear matter. We point out that the large variety 
and abundance of produced hypernuclei is an important 
advantage of relativistic heavy ion collisions in comparison with the 
traditional hypernuclear methods concentrated on reactions leading 
only to a few species. 

Previously nucleosynthesis was under examination mostly for stable nuclei. 
Now we reach the next stage were also unstable conventional nuclei (e.g., $^5$Li, 
$^8$Be), single and double hypernuclei can be produced in the same reaction 
events. We suggest that some problems of the hypernuclear physics can be 
resolved by involving these reactions in which there is a great yield of 
exotic nuclei, e.g., H-dibaryons. In addition, we predict 
a considerable production of $^4_{\Lambda \Lambda}$H hypernuclei, which 
were already observed in other hypernuclear experiments. The production rate 
of these nuclei is high enough for their detection. 
By comparing the yields of different hypernuclei one can extract 
important information about hypermatter in which the nuclei 
are formed, and about the properties of exotic hypernuclei \cite{Buy23,Buy18}. 
This information is also of utmost importance in astrophysics for models 
describing stellar matter in supernova explosions and in binary neutron 
star mergers. Big hypernuclear isotopes and double hypernuclei can be 
singled out from large pieces of hypernuclear matter in local chemical 
equilibrium. Therefore, their formation process can provide complementary 
information on the matter, and a deeper knowledge on the hyper-nucleosynthesis at 
low densities will be obtained. The correlation measurements in the 
experiments and the comparison of different nuclei/hypernuclei yields in 
same reactions are a straightforward method to obtain this new information. 

The authors acknowledges German Academic Exchange Service (DAAD) 
support from a PPP exchange grant and the Scientific and 
Technological Research Council of  T\"urkiye (TUBITAK) support 
under Project No. 121N420.  
T.R. acknowledges support through the Main-Campus-Doctus fellowship 
provided by the Stiftung Polytechnische Gesellschaft Frankfurt am Main 
and further thanks the Samson AG for their support. 
N.B. thanks J.W. Goethe University Frankfurt am Main for hospitality. 
Computational 
resources were provided by the Center for Scientific Computing (CSC) of 
the Goethe University and the "Green Cube" at GSI, Darmstadt. 

\end{document}